\documentclass[%
 reprint,
 amsmath,amssymb,
]{revtex4-2}

\usepackage[T1]{fontenc}

\usepackage{graphicx}
\usepackage{float}
\newfloat{scheme}{htbp}{los}
\floatname{scheme}{Scheme}
\floatname{chart}{Chart}
\newfloat{graph}{htbp}{loh}

\usepackage{xcolor}
\usepackage{lineno}

\usepackage{hyperref}
\hypersetup{
    colorlinks=true,
    linkcolor=blue,
    filecolor=magenta,      
    urlcolor=cyan,
    }

\begin{document}

\title{Tailoring Attosecond Charge Migration in Native Molecular Ions}

\author{Evan Munaro-Langloÿs}
\email{evan.langloys@univ-lyon1.fr}
\author{Franck Lépine}
\author{Victor Despré}
\email{victor.despre@univ-lyon1.fr}
\affiliation{%
	Universite Claude Bernard Lyon 1, CNRS, Institut Lumière Matière, F-69622
	Villeurbanne, France
}%

\date{\today}

\begin{abstract}
	Attosecond chemistry involves developing strategies to manipulate electronic
	coherent waves in molecules, which can influence the outcome of photoinduced
	reactions. While recent progress in this field calls for investigations of
	increasingly complex isolated or embedded systems, theoretical predictions on
	attosecond charge migration have remained limited to native neutral species.
	Since molecules in nature often carry a native charge, there is potential
	biological and chemical interest in determining whether attosecond charge
	migration is affected by an additional charge. In this study, we employ
	high-level correlated methods to study purely electronic dynamics induced by
	hole-mixing in molecular ions. Our results, obtained for a series of neutral,
	protonated and deprotonated molecules, reveal that the likelihood of observing
	attosecond electron dynamics can either be degraded or improved by the
	presence of an initial charge, and that the existence of the dynamics
	is correlated with the strength of electron correlation. These findings will
	stimulate further experimental and theoretical investigations into this
	unexplored field of attosecond dynamics in molecular ions.
\end{abstract}

\maketitle

\section{Introduction}

Attosecond science is an exciting research field that offers new ways to study
molecular systems with unparalleled temporal and spatial resolution
\cite{Corkum2007,Krausz2009}. It also offers new control strategies by directly
affecting electronic degrees of freedom, as proposed in attochemistry
\cite{Calegari2023}. According to this concept, an attosecond pulse creates a
time-dependent coherent superposition of electronic states corresponding to an
ultrafast electronic wave that migrates through the molecular backbone without
the need of nuclear motion \cite{calegari2014ultrafast}. This process, known as
‘charge migration’ \cite{Kuleff2014}, can impact a chemical reaction
\cite{Remacle1998}. Theoretical
\cite{Despre2015,mignolet2014charge,lara-astiaso2016decoherence,vacher2017electron,yuan2019ultrafast,folorunso2021molecular,zhang2024cavity,tremblay2026persistent,scheidegger2025can}
and experimental
\cite{calegari2014ultrafast,lara-astiaso2018attosecond,kraus2015measurement,matselyukh2022decoherence,schwickert2022electronic,driver2024Coherent}
investigations of attosecond charge migration have been conducted in the
context of bound-bound excitation and following ionisation in neutral
molecules. While attosecond charge migration can be produced by coherent
superposition of electronic states in general, attosecond hole migration
produced by ionization is especially interesting as it allows to consider the
case where the dynamics is solely driven by electron correlation
\cite{Cederbaum1999}. In weakly correlated outer-shell electrons,  the removal
of an electron from the highest occupied Hartree–Fock molecular orbitals
(HOMOs) does not induce significant hole dynamics driven by correlation. This
arises from Koopmans’ theorem, which predicts that when electron correlation
effects are negligible, one-hole ({\it 1h}) configuration is a good
approximation of the stationary states of the cation. However, in some
molecules, outer-shell electrons are strongly correlated and Koopmans’ theorem
breaks down. This occurs when 1h configurations are coupled through electron
correlation effects, more precisely via excited 2h1p configurations of the
system \cite{Luennemann2008a}. This situation is commonly referred to as hole
mixing \cite{Niessen1982}. Studying attosecond hole migration in the case of
hole mixing is a perfect playground to investigate correlation effects
\cite{Sansone2012,Kraus2018}.

So far, attosecond charge migration investigations have focused on simple,
isolated neutral molecules. However, improvements in experimental techniques
now allow us to study increasingly complex systems in the gas
\cite{herve2021ultrafast}, liquid \cite{li2024attosecond} and condensed phases
\cite{cavaletto2025attoscience} that have potential chemical and/or biological
significance \cite{Calegari2016}. However, while the focus has been on neutral
molecules, molecules in nature often carry a charge that significantly impacts
their properties. This limits the predictability of current investigations
concerning their chemical or biological relevance. Therefore, a natural
question is to ask whether attosecond charge migration survives in charged
molecules.

In this article, we examine situations in which hole migration is initiated in
molecules to which a proton has been added or removed (i.e. protonated and
deprotonated molecules). This corresponds, respectively, to positively charged
(cation) and negatively charged (anion) species. Besides the fact that
deprotonated and protonated systems are common in nature, studying neutral,
protonated and deprotonated molecules allows to compare molecules while
maintaining the total number of electrons when changing the charge. Considering
attosecond migration in molecular hole-mixing states, we demonstrate that
protonation and deprotonation can suppress or accelerate attosecond dynamics.
Our investigation of a series of molecules reveals a general trend linking
attosecond dynamics, initial charge localization and electron correlation
strength. Notably, these systems can also be accessed experimentally in an
isolated or liquid phase, making this work relevant to future experimental
investigations.

\section{Results}

\begin{figure*}
	\begin{center}
		\includegraphics[width=\textwidth]{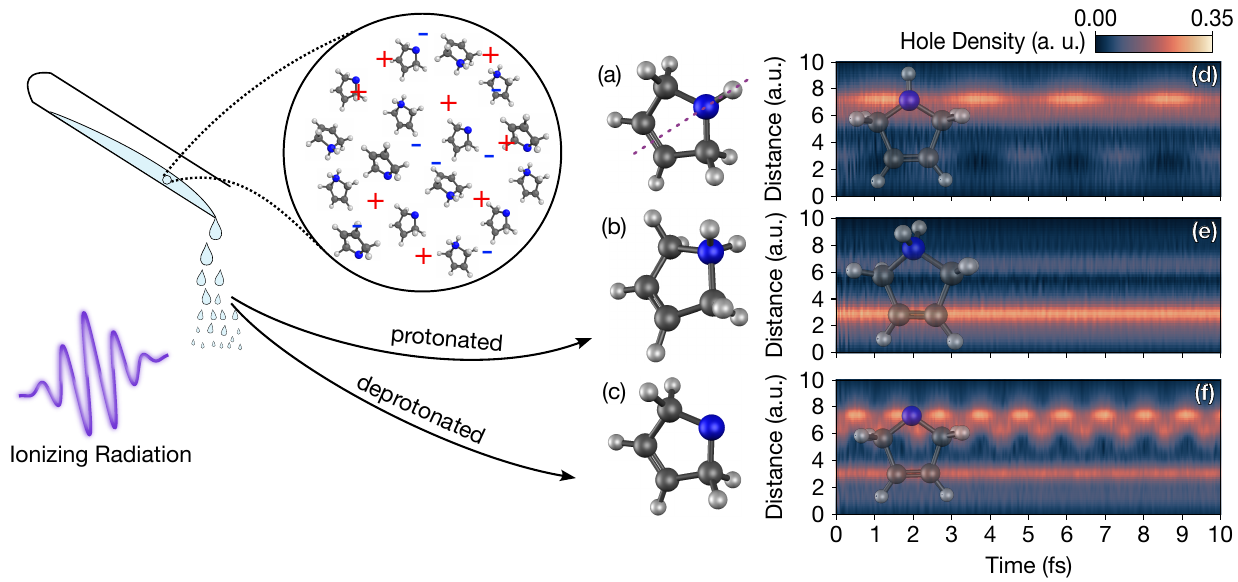}
	\end{center}
	\caption{Hole migration following ionization of an outer-shell electron in neutral, protonated, and deprotonated 3-pyrroline. Panels (a), (b), and (c) show ball-and-stick representations of the neutral, protonated, and deprotonated geometries, respectively, optimized at the MP2/cc-pVDZ level. Panels (d), (e), and (f) show the time evolution of the hole density following ionization of an outer-shell electron for the neutral, protonated, and deprotonated species, respectively. We project the hole density onto the axis passing through the nitrogen atom and the center of the double bond adjacent to the nitrogen, as indicated in the neutral geometry in panel (a). For the neutral and protonated species, the initial state corresponds to the (HOMO)$^{-1}$ configuration. For the deprotonated species, the initial state is a coherent 50\%–50\% superposition of the (HOMO)$^{-1}$ and (HOMO-2)$^{-1}$ configurations.}
	\label{fig:first_panel}
\end{figure*}

To study the role of electron correlation in hole migration in ionic molecules,
we consider molecules that exhibit hole mixing in their outermost electronic
structure. These systems are known to exhibit pronounced charge migration.
Furthermore, most of the selected neutral systems have been shown to exhibit
long-lived electronic coherence, making them promising candidates for future
experimental research \cite{Despre2018,Scheidegger2022}. First, we present the
results for the 3-pyrroline molecule, and then we discuss our conclusions based
on the study of a series of molecules.

First, we consider the protonation of 3-pyrroline, which involves attaching a
proton to the nitrogen atom. We also consider the deprotonation of 3-pyrroline,
which involves removing the proton bonded to the same nitrogen atom (see
Figs.~\ref{fig:first_panel}a, \ref{fig:first_panel}b, and
\ref{fig:first_panel}c). Although protonation and deprotonation can also occur
at the double bond, these processes significantly alter the molecular geometry.
Such structural changes make direct comparisons between the electronic
structures of the neutral molecule and its corresponding ions difficult, so
these cases are not considered here.  More generally, throughout this work, we
restrict our analysis to protonated or deprotonated species whose geometric
structures remain sufficiently close to those of the neutral molecules, in
order to focus our analysis on the added/removed charge and the effect of
electron correlation. For all molecules studied, we follow the time evolution
of the hole density after the ionisation of an outer-shell electron
\cite{Breidbach2003}. The hole density is defined as the difference between the
electron density of the stationary molecular ground state and that of the
time-dependent photoionized state at time t, and time propagation is performed
using the multi-electronic wave-packet propagation method \cite{Kuleff2005},
with a Hamiltonian constructed within the non-Dyson algebraic diagrammatic
construction (nD-ADC(3)) framework \cite{Schirmer1998,Breidbach2007}.
Additional methodological details and references are provided in the Methods
section.

In the neutral 3-pyrroline molecule, the (HOMO)$^{-1}$ and (HOMO$-1$)$^{-1}$
configurations exhibit a hole-mixing structure. In practice, this means that
preparing the molecule in one of these configurations—for instance,
(HOMO)$^{-1}$, leads to a coherent superposition of the two cationic
eigenstates involved in the hole mixing, namely the first and second ionic
states (see Fig.~\ref{fig:spec_prot}a), which are mixtures of the (HOMO)$^{-1}$
and (HOMO$-1$)$^{-1}$ configurations. This occurs because neither the
(HOMO)$^{-1}$ nor the (HOMO$-1$)$^{-1}$ configuration represents an approximate
stationary state of the ionized molecule. Such a preparation enables the charge
to migrate between different sites along the molecular backbone.

In general, the (HOMO)$^{-1}$ and (HOMO$-1$)$^{-1}$ configurations correspond
to different expectation values of the electron-density operator. Consequently,
the time evolution of the configuration populations induced by electron
correlation can result in a redistribution of the molecular charge density,
referred to as correlation-driven charge migration arising from hole mixing. In
3-pyrroline, removal of an electron from the HOMO initiates a migration of the
created hole between the nitrogen atom and the double bond in front of it (see
Fig.~\ref{fig:first_panel}d). The period of this dynamics is approximately
2.5~fs, consistent with the energy separation of 1.68~eV between the first and
second ionized states.

\subsection{Protonation to Control the Position \\of the Hole}

\begin{figure}
	\begin{center}
		\includegraphics[width=0.5\textwidth]{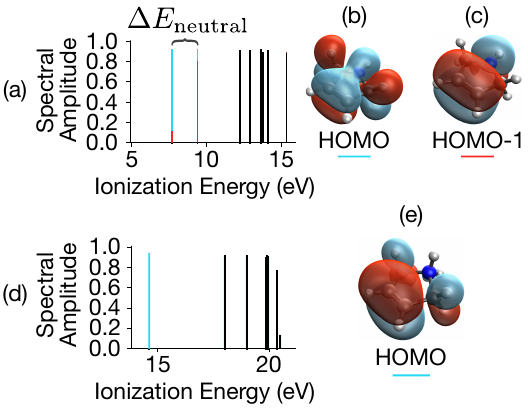}
	\end{center}
	\caption{Effect of the protonation on the energy spectra of the 3-pyrroline molecule. (a) and (b) are the energy spectra of the neutral and protonated 3-pyrroline molecule computed at the nD-ADC(3)/cc-pVDZ level. The colors of the lines indicate the contributions from the different {\it 1h} configurations to the energy eigenstates. The Hartree-Fock orbitals corresponding to the {\it 1h} configurations are shown on the right of the spectra.}
	\label{fig:spec_prot}
\end{figure}

To investigate how protonation affects hole migration, we compute the nD-ADC(3)
energy spectrum of protonated 3-pyrroline (see Fig.~\ref{fig:spec_prot}d).
First, we observe that protonation leads to a substantial increase in the
ionization potential, from 7.7~eV in the neutral molecule to 14.7~eV in
protonated 3-pyrroline. This positive shift in the ionization potential upon
addition of a positive charge is expected and has previously been discussed in
the context of intermolecular Coulombic decay \cite{Kryzhevoi2011,Kumar2022}.
It can be understood from the fact that the additional positive charge
strengthens the attractive potential acting on the electrons, thereby
increasing the energy required for ionization.

Comparing the lowest ionized states of neutral (Fig.~\ref{fig:spec_prot}a) and
protonated (Fig.~\ref{fig:spec_prot}d) 3-pyrroline reveals that the
outer-valence electronic structure of the protonated cation differs markedly
from that of the neutral molecule after ionization. Of the two states
responsible for hole mixing in the neutral system, only one remains in the
protonated case, and the hole-mixing character disappears. This remaining state
is well separated from higher-lying states by an energy gap of approximately
3.5~eV and is dominated by a single {\it 1h} configuration, namely
(HOMO)$^{-1}$. Visualization of the HOMO of protonated 3-pyrroline shows that
it corresponds to an orbital largely localized on the double bond, opposite to
the added positive charge (see Fig.~\ref{fig:spec_prot}e). These results
indicate that, in the protonated molecule, electron removal is energetically
favored from the double bond rather than from other regions of the molecule.
This contrasts with the neutral 3-pyrroline case, where the orbitals involved
in the hole mixing structure (see Figs.~\ref{fig:spec_prot}b and
\ref{fig:spec_prot}c), demonstrate an energetic preference for charge
delocalization over the molecular framework.

For the protonated system, removal of an electron from the HOMO does not
trigger any appreciable ultrafast dynamics. Instead, the hole initially created
on the double bond remains localized throughout the time propagation (see
Fig.~\ref{fig:first_panel}e). Notably, the hole in the electronic density
produced by HOMO ionization stays spatially separated from the added positively
charged proton. Interpreting the hole as an effective positive charge, this
behavior resembles the classical picture of electrostatic repulsion between two
positive charges. We emphasize, however, that the apparent stationarity of the
hole in our simulations does not imply indefinite localization. Ionization of
the HOMO may still initiate charge-transfer processes driven by nuclear motion,
which occur on longer timescales—typically tens of femtoseconds to picoseconds.
Finally, the stationary character of the hole can already be anticipated from
the nD-ADC(3) spectrum in Fig.~\ref{fig:spec_prot}d, which shows that the
(HOMO)$^{-1}$ configuration closely approximates a stationary state of the
ionized molecule.

The effect observed here for 3-pyrroline is in fact general across all
molecules investigated, 3-pyrroline, 2,5-dihydrofuran, propiolic acid,
pent-4-enal, but-3-ynal, PENNA, MePeNNA, and propynamide (see Supplementary
Materials). The consistent spectral modification observed across chemically
diverse systems indicates that protonation acts as a strong perturbation of the
molecular potential, largely independent of molecular geometry or atomic
composition.

\subsection{Deprotonation to Control the Timescale\\of the Migration}

\begin{figure*}[t]
	\begin{center}
		\includegraphics[width=\textwidth]{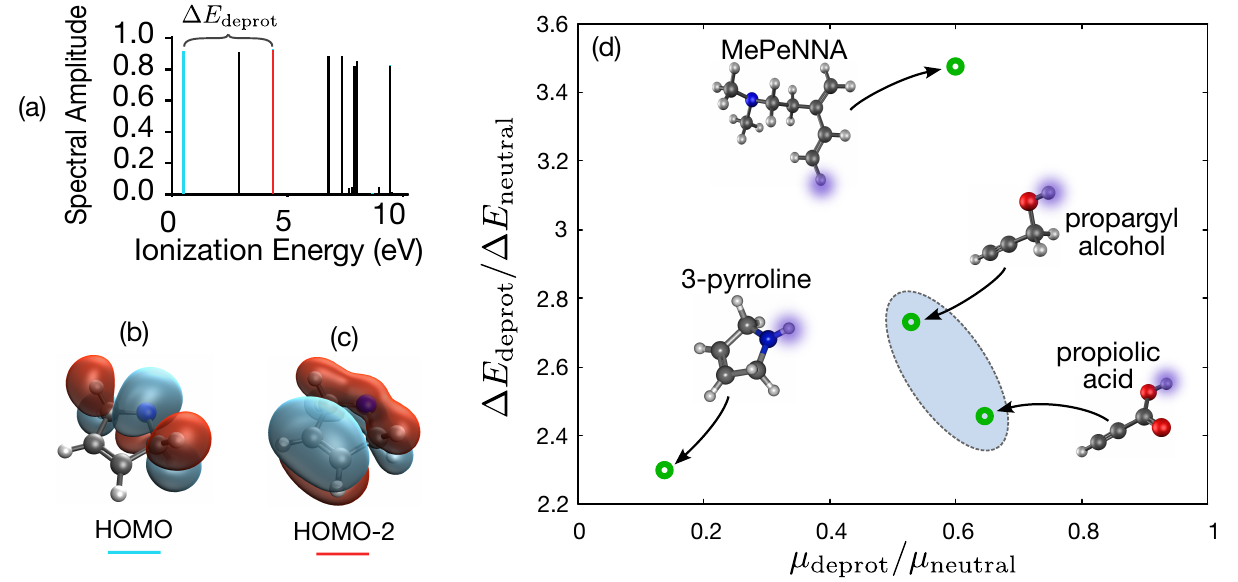}
	\end{center}
	\caption{\small Effect of the deprotonation on the electronic structure of
		strongly correlated molecules. (a) Energy spectrum of the deprotonated
		3-pyrroline molecule computed at the nD-ADC(3)/cc-pVDZ level. The colors of
		the lines indicate the contributions from the different {\it 1h}
		configurations to the energy eigenstates. The Hartree-Fock orbitals
		corresponding to the {\it 1h} configurations are shown under the spectra. (d)
		Evolution of the migration timescale plotted as a function of the mixing
		evolution for some molecules having a hole mixing in their first states. Each
		point corresponds to one molecule for which the geometry is shown as an inset.
		The blue highlight on the molecular geometries indicate which proton has been
		removed to produce the deprotonated species.}\label{fig:deprot_panel}
\end{figure*}

We now present our results for deprotonated molecules. The nD-ADC(3) spectrum
of deprotonated 3-pyrroline shows that deprotonation affects the electronic
structure in a markedly different way from protonation (see
Fig.~\ref{fig:deprot_panel}a). First, deprotonation considerably lowers the
ionization potential, decreasing from 7.7~eV in the neutral molecule to 0.51~eV
in deprotonated 3-pyrroline. The interpretation of this global negative energy
shift parallels that of the positive shift observed upon protonation: removing
a proton reduces the attractive potential acting on the electrons, thereby
lowering the energy required to ionize the molecule.

The first ionized state of deprotonated 3-pyrroline is dominated by a single
	{\it 1h} configuration with a hole in the HOMO. A small contribution from the
(HOMO$-2$)$^{-1}$ configuration remains, indicating that although strongly
reduced, the hole-mixing structure survives deprotonation. Visualization of the
HOMO shows that this orbital is primarily localized on the nitrogen atom from
which the proton has been removed (see Fig.~\ref{fig:deprot_panel}b). Following
the same reasoning as in the protonation case, we conclude that deprotonation
of the nitrogen atom makes electrons in its vicinity significantly easier to
ionize.

Up to this point, the effects of protonation and deprotonation on the
electronic structure appear similar, aside from the obvious differences arising
from the opposite signs of the introduced charges. However, in the deprotonated
case, two ionized states appear that are clearly separated from the denser
manifold of states beginning above 7.5~eV. The second and third ionized states
are dominated by the (HOMO$-1$)$^{-1}$ and (HOMO$-2$)$^{-1}$ configurations,
respectively. The HOMO$-1$ belongs to a different irreducible representation
than the HOMO and HOMO$-2$. Because molecules can be aligned experimentally,
states of different symmetry can be selectively addressed using linearly
polarized light \cite{he2022filming}. For this reason, the second ionized state
is not considered further in the following discussion.

Owing to the weak residual hole mixing, ionization from either the HOMO or
HOMO$-2$ alone is expected to generate only weak dynamics, since essentially a
single ionic state would be populated. To assess whether the charge migration
observed in the neutral molecule is preserved, we instead prepare a wave packet
corresponding to equal ionization amplitudes from the two orbitals, thereby
creating a significant coherent superposition of both states. Such a situation
is realistic when deprotonated 3-pyrroline is ionized by a sufficiently short
pulse. The resulting dynamics, shown in Fig.~\ref{fig:first_panel}f, reveal
that similar charge migration than for the neutral occurs on a substantially
shorter timescale. This directly reflects the increased energy separation
between the states forming the wave packet compared to the neutral case.

Overall, the main impact of deprotonation appears twofold: a weakening of hole
mixing--signaling a reduction of electron-correlation effects--and a
corresponding shortening in the timescale of the charge-migration dynamics. To
quantify the reduction of mixing induced by deprotonation and to enable
comparisons across different systems, we introduce the mixing quantity, defined
as follows. Consider an ionic state expressed as a superposition of two {\it
		1h} configurations with complex amplitudes $x_1$ and $x_2$, as typically occurs
in the presence of hole mixing. The mixing is defined as
\begin{equation}
	\mu = \mathrm{max}\left(\frac{|x_1|}{|x_2|},\frac{|x_2|}{|x_1|}\right).
\end{equation}
A value of one corresponds to an equal 50\%–50\% superposition of the two
configurations, whereas a value approaching zero indicates a nearly pure {\it
		1h} configuration. For 3-pyrroline, the mixing of the first ionized state
decreases from 0.37 in the neutral molecule to 0.05 in the deprotonated
species, demonstrating that deprotonation leads to a strong reduction of hole
mixing.

It is instructive to examine how deprotonation affects other molecules
exhibiting hole mixing in their lowest ionized states, namely MePeNNA,
propiolic acid, and propargyl alcohol. For each system, we compute the
nD-ADC(3) spectra of both neutral and deprotonated molecules and analyze the
evolution of hole mixing. The results are summarized in
Fig.~\ref{fig:deprot_panel}d, where the ratios of charge-migration timescales
between neutral and deprotonated species are plotted as a function of the
corresponding ratios of the mixing parameters. The full spectra are provided in
the Supplementary Information.

For all molecules studied, the mixing decreases upon deprotonation, as
evidenced by ratios consistently smaller than unity. Among the investigated
systems, 3-pyrroline exhibits the strongest reduction in mixing. In addition,
the charge-migration timescale becomes at least twice as fast for all molecules
considered. Interestingly, propiolic acid and propargyl alcohol—two molecules
with closely related structures—cluster in the same region of the plot,
suggesting a structural influence on the quantitative response to
deprotonation.

\section{Discussions}

While our results show that protonation and deprotonation affect outer-valence
hole-mixing structures, and the resulting hole migration, in markedly different
ways, it is striking that the underlying physical origin of these differences
can be traced back to the same simple principle. Protonation modifies the
spatial localization of the sites from which electrons are most easily ionized.
More precisely, the atom to which the proton is attached becomes harder to
ionize than the rest of the molecule. This can be understood straightforwardly:
the added proton locally increases the attractive potential experienced by the
electrons, so electrons located near the proton require more energy to be
removed than those farther away.

At the level of hole mixing, this effect causes one of the two mixed {\it 1h}
configurations present in the neutral molecule to disappear. Protonation
therefore suppresses hole mixing and replaces it with a single dominant {\it
		1h} configuration characterized by a hole localized away from the added proton.
The observation that similar spectral modifications occur across chemically
diverse molecules indicates that protonation acts as a strong perturbation of
the molecular potential, largely independent of molecular geometry or atomic
composition.

In contrast, deprotonation generally preserves the hole-mixing structure of the
neutral molecule. As in the protonation case, deprotonation alters the relative
ionization propensity of different sites, making electron removal easier near
the deprotonation site. However, the overall negative charge introduced by
deprotonation promotes greater electronic delocalization. Consequently,
although the preferred localization of the hole after ionization is modified,
it remains distributed over the molecular framework rather than confined to a
single region. The enhanced ionizability near the deprotonated site nonetheless
breaks the equivalence between previously mixed configurations, thereby
reducing the strength of hole mixing.

The contrasting behaviors can therefore be rationalized by the nature of the
added charge. In protonated molecules, the additional positive charge
corresponds to a proton with a well-defined spatial position. In deprotonated
molecules, by contrast, the extra negative charge is an electron that can
delocalize throughout the molecular system.

Although simple electrostatic arguments explain much of the observed
robustness, it is notable that strong correlation effects can survive
deprotonation, making it a promising approach for studying electron correlation
among outer-valence electrons in molecular systems. To explore this aspect, we
compared the hole mixing of neutral and deprotonated species using the mixing
$\mu$ introduced above, which quantifies the coupling between configurations
involved in hole mixing. Because this coupling originates from electron
correlation, the mixing can be regarded as an indicator of correlation strength
among outer-valence electrons. The systematic reduction of the mixing observed
for all studied molecules indicates that deprotonation generally weakens
electron correlation. This trend is expected, as outer-valence electrons in
anions are, on average, located farther from the molecular core, reducing their
mutual interactions. In addition, by making the deprotonation site easier to
ionize, deprotonation breaks the energetic equivalence between previously mixed
ionization channels, further diminishing configuration mixing.

The magnitude of the deprotonation effect varies among the molecules
considered. For example, deprotonation at the nitrogen atom of 3-pyrroline
almost completely suppresses hole mixing, whereas in the other systems the
reduction remains below a factor of two. A key feature distinguishing
3-pyrroline is its ring structure, which causes outer-valence electrons to be
strongly influenced by local electrostatic changes throughout the entire ring,
leading to a pronounced reduction of mixing. A second outlier is MePeNNA, which
exhibits a particularly strong shortening of the charge-migration period. In
this case, the larger molecular size appears to play an important role: the
impact of deprotonation is more localized, allowing the two mixed states,
because of their different spatial localizations, to be affected unequally,
thereby increasing their energy separation. The strong molecule-dependent
response suggests that the effect of deprotonation on hole mixing is governed
by how the additional negative charge redistributes across the molecular
framework.

Finally, we highlight several experimental perspectives emerging from these
results. A particularly attractive consequence of deprotonation is the
substantial lowering of the ionization potential to only a few electron-volts,
opening new experimental pathways for initiating charge migration. For
instance, correlation-driven charge migration arising from hole mixing has
recently been shown to be accessible via multiphoton ionization using infrared
light \cite{GuiotduDoignon2025}. A reduced ionization potential implies that
lower infrared intensities are sufficient, enabling charge migration to be
triggered with femtosecond infrared pulses of moderate intensity. The infrared
pulse can initiate the dynamics, while an XUV attosecond pulse generated by
high-harmonic generation can probe the ensuing electronic motion.
Alternatively, the low ionization energies of the hole-mixing states make it
possible to trigger charge migration using short pulses in the visible or
ultraviolet spectral ranges, enabling UV pump–XUV probe schemes for resolving
ultrafast charge dynamics \cite{Wanie2024}.

We have demonstrated how the addition or removal of charge through protonation
or deprotonation affects charge migration arising from the strongly correlated
hole-mixing regime. In the case of protonation, correlation effects are largely
suppressed and no significant electronic dynamics are observed. In contrast,
deprotonation preserves correlation effects, although in a reduced form, and
leads to faster charge-migration dynamics.

These results represent an important first step toward extending the use of
attosecond technologies to new research areas, most notably chemistry and
biology, where charged systems are ubiquitous. One of the most promising
aspects of attosecond science is its ability to reveal mechanisms governed by
electron correlation. By clarifying how the addition or removal of charge
influences a cornerstone mechanism of attosecond dynamics, our work opens new
opportunities for exploring correlated electronic motion in complex molecular
environments.

A central open question concerns how quantum electronic effects influence
chemical reactivity and biological function, in other words, how ultrafast
quantum phenomena can propagate toward longer time and length scales. Our
results provide an initial step in this direction. Extending the present study
to systems with different structural motifs and to other correlation regimes,
such as satellite states, would further deepen our understanding of the
observed mechanisms. Given the fundamental importance of charged species in
solution, extending these investigations toward molecules interacting with
solvent \cite{zhou2025state} or even in liquid phase
\cite{moore2025solvation,zhang2025intermolecular} also appears particularly
timely. We hope that our findings will stimulate future theoretical and
experimental studies along these lines.

\section{Methods}

For all molecules and ions, we optimized the ground-state geometry at the
MP2/cc-pVDZ level using the Gaussian package \cite{g16}. We calculated
restricted Hartree–Fock orbitals with the GAMESS-UK package using the cc-pVDZ
basis set \cite{Wilson1999}. For negative ions, we performed unrestricted
calculations with a diffuse basis set to assess the validity of the restricted
treatment. For all anions considered, the UHF/aug-cc-pVDZ and RHF/cc-pVDZ
approaches yield essentially identical orbitals, confirming the validity of the
restricted description for these systems.

We construct the electronic Hamiltonian of the ionized molecules at their
ground-state geometries using the non-Dyson third-order algebraic diagrammatic
construction [nD-ADC(3)] method \cite{Schirmer1998}. In this approximation, we
represent the Hamiltonian in a basis of correlated configurations ({\it 1h},
{\it 2h1p}, \ldots) built upon the exact electronic ground state of the
molecule \cite{Mertins1996,Schirmer2004}. To obtain explicit matrix elements in
this basis, the exact ground state is approximated to a given order in
perturbation theory; the $n$th order defines the nD-ADC($n$) approximation. At
third order, the nD-ADC scheme fully accounts for the coupling between {\it 1h}
and {\it 2h1p} configurations and thus accurately describes hole-mixing
structures.

We briefly describe the treatment of charge migration in this work. The first
issue in post-ionization dynamics is the choice of the initial state for
propagation. In this study, we select initial states as pure {\it 1h}
configurations with a hole in a Hartree–Fock orbital. This choice highlights
the role of electronic correlation in the dynamics. According to Koopmans'
theorem, {\it 1h} configurations are energy eigenstates of the ionized system;
therefore, preparing the system in such a state should not induce dynamics if
correlations are negligible. Consequently, any observed dynamics originating
from such an initial state can be attributed to electron interactions beyond
the Hartree–Fock level, that is, electronic correlation. For an explicit
description of the probe pulse, see Ref. \cite{Dey2022}.

After defining the initial state, we perform time propagation using a short
iterative Lanczos scheme \cite{Kuleff2005}. To obtain the state at time $t +
	\Delta t$, we apply the time-evolution operator for a time step $\Delta t$ to
the state at time $t$ within a suitably chosen reduced subspace, namely the
Krylov space. An important property of this scheme is that we can express the
Hamiltonian defining the evolution operator in any basis; thus, we do not need
to diagonalize the nD-ADC Hamiltonian to obtain the time evolution. Another
advantage of this method is that it provides explicit control over the error at
each time step by adjusting the time-step duration and the dimension of the
Krylov space. After propagation, we obtain an explicit representation of the
electronic state vector at each time step, which enables calculation of
expectation values of observables such as the hole density.

\section*{Acknowledgements}

The authors thank Alexander Kuleff for fruitful discussions.

\section*{Authors Contributions}
F.L. and V.D. conceived the project. E.M.L. performed the simulations. All the
authors contributed equally to the manuscript.

\end{document}


\title{Supplementary Information: Tailoring Attosecond Charge Migration in Native Molecular Ions}

\author{Evan Munaro-Langloÿs}
\author{Franck Lépine}
\author{Victor Despré}
\affiliation{%
	Universite Claude Bernard Lyon 1, CNRS, Institut Lumière Matière, F-69622
	Villeurbanne, France
}%

\maketitle

\section{Protonation}
\begin{figure}
    \centering
    \includegraphics{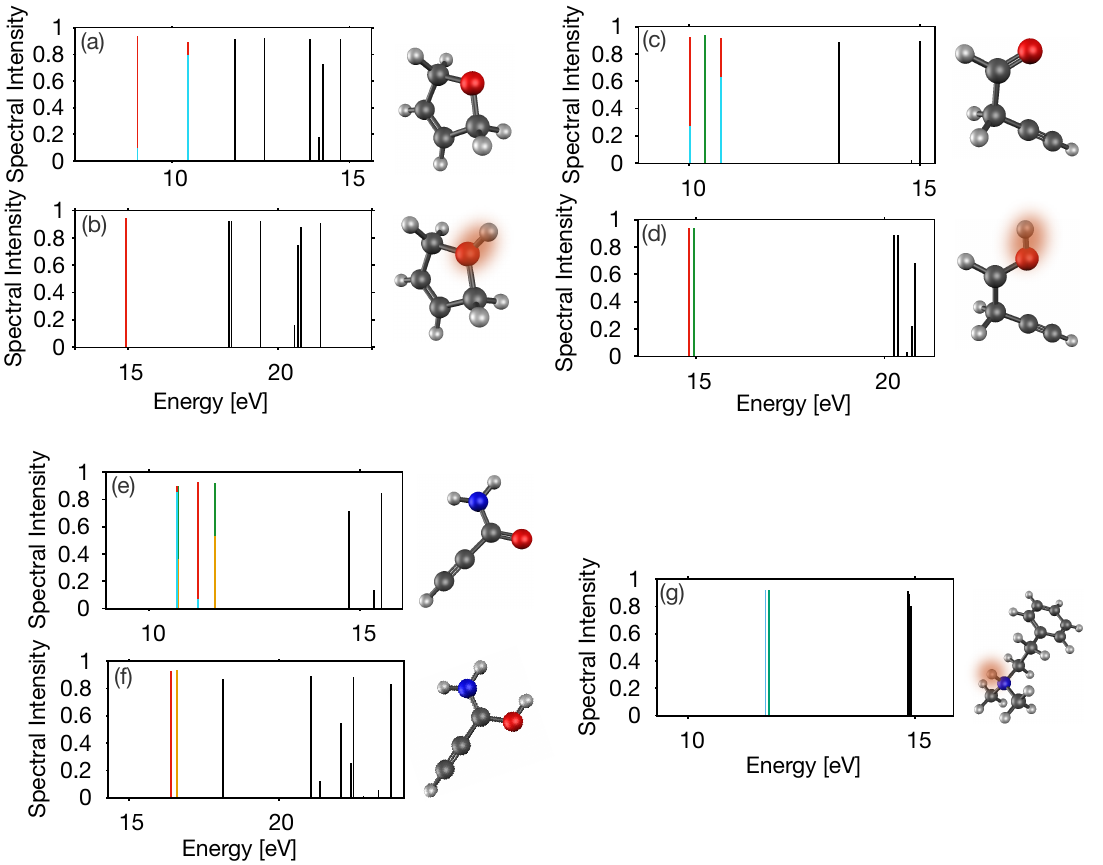}
    \caption{For each molecule, the nD-ADC(3)/cc-pVDZ spectra of the neutral and protonated species are shown. The systems considered are (a,b) 2,5-dihydrofuran, (c,d) but-3-ynal, (e,f) propynamide, and (g) PENNA. Each line represents an ionized state and is positioned according to its ionization energy. The line colors indicate the contributions of different one-hole (1h) configurations to each state, while states not discussed in the text are shown in black. Ball-and-stick representations of the optimized geometries, calculated at the MP2/cc-pVDZ level, are displayed to the right of each spectrum. The red highlight indicates the position of the added proton.}
    \label{fig:supp1}
\end{figure}

In addition to 3-pyrroline discussed in the main text, we investigated the effect of protonation in
several molecules that exhibit hole mixing in their first ionized states. Although all systems follow
the general behavior described in the main text, some display specific features. The results are
summarized in two figures: Figure \ref{fig:supp1} shows molecules for which only protonation was examined,
whereas Figure \ref{fig:supp2} presents systems investigated under both deprotonation and, where possible,
protonation.

First, 2,5-dihydrofuran, a cyclic molecule structurally similar to 3-pyrroline, exhibits the same
response to protonation. The well-isolated hole mixing observed in the neutral species (Fig. \ref{fig:supp1}a) is
replaced by a single one-hole (1h) configuration upon protonation of the oxygen atom (Fig. \ref{fig:supp1}b).
Likewise, protonation of the nitrogen atom in the larger MePeNNA molecule leads to the same
behavior (Fig. \ref{fig:supp2}d, f).

The PENNA molecule differs in that its neutral form exhibits triple hole mixing involving the (HOMO-1)$^{-1}$ , (HOMO-2)$^{-1}$, and (HOMO-3)$^{-1}$ configurations (see Ref. 22 in the main text). Upon protonation of the nitrogen atom, this mixing disappears and is replaced by two distinct 1h states
that are close in energy (Fig. \ref{fig:supp1}g). Analysis of the contributing orbitals shows that both resulting
configurations localize the hole far from the added proton, supporting the same interpretation
proposed for the two-state case.

But-3-ynal and propiolic acid differ from 3-pyrroline by the presence of additional 1h states within
the energy window of the hole mixing. Among these systems, but-3-ynal has the simplest structure.
In the neutral species, one state lies between the two mixed states (Fig. \ref{fig:supp1}c). After protonation, the
hole mixing collapses into a single 1h state, while the intermediate state remains (Fig. \ref{fig:supp1}d). In
propiolic acid, the neutral species contains two additional states besides the hole-mixed pair (Fig.
\ref{fig:supp2}a). Upon protonation, the hole mixing is replaced by a single 1h state, and only one of the
additional states persists (Fig. \ref{fig:supp2}c). The two remaining states become nearly degenerate in the
protonated species, reflecting the increased symmetry induced by protonation.

Finally, neutral propynamide exhibits two independent hole-mixing structures within the same
energy range (Fig. \ref{fig:supp1}e). Following protonation, both mixings reduce to single 1h configurations, such
that the two lowest ionized states of protonated propynamide correspond to pure 1h states (Fig.
\ref{fig:supp1}f).

\section{Deprotonation}
\begin{figure}
    \centering
    \includegraphics{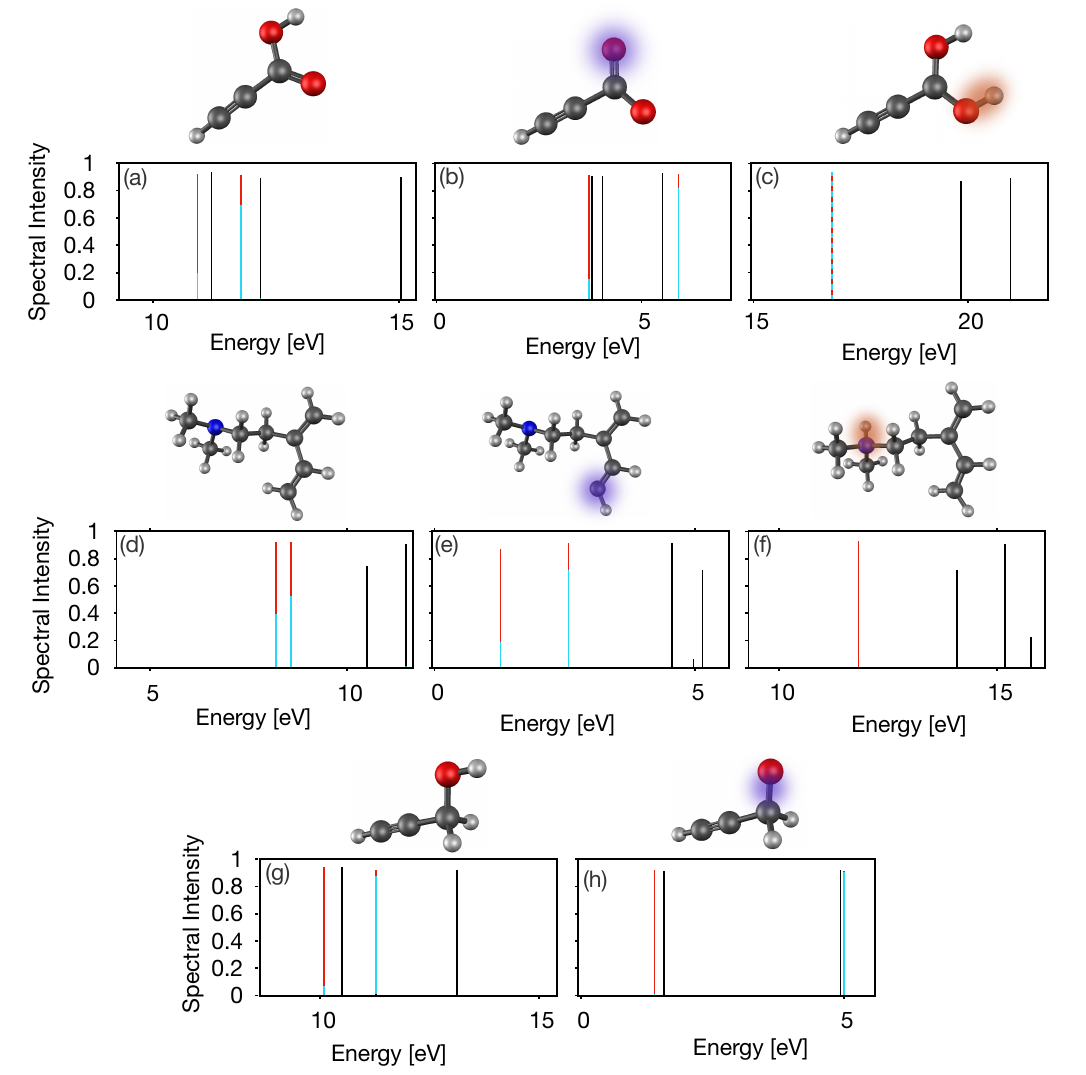}
    \caption{For each molecule, the nD-ADC(3)/cc-pVDZ spectra of the neutral, deprotonated, and protonated species are shown. See Fig. 1 for the interpretation of the spectra. The systems considered are (a–c) propiolic acid, (d–f) MePeNNA, and (g,h) propargyl alcohol, for which no stable geometry is found in the protonated form. Ball-and-stick representations of the optimized geometries, calculated at the MP2/cc-pVDZ level, are displayed above each spectrum. In protonated molecules, a red highlight indicates the position of the added proton, whereas in deprotonated molecules a blue highlight marks the atom from which the proton has been removed.}
    \label{fig:supp2}
\end{figure}

Figures \ref{fig:supp2}b, \ref{fig:supp2}e, and \ref{fig:supp2}h present the nD-ADC(3) spectra of deprotonated propiolic acid, MePeNNA,
and propargyl alcohol, respectively. As discussed in the main text, all three molecules retain their
hole-mixing structure upon deprotonation, although both the energy spacing between states and
the degree of configurational mixing are modified.

\newpage

\newpage

